\documentclass[12pt]{iopart}
\begin{document}

\title[A Doppler-looking redshift observed in the lab; application to quasars]
{A Doppler-looking redshift observed in the labs; application to quasars}

\author{J Moret-Bailly}

\address{ LPUB, Université de Bourgogne BP 400, 21011 Dijon, France

jmb@jupiter.u-bourgogne.fr}

\begin{abstract}
It is known since 1968 that the interaction of a pulse of light with matter
redshifts the spectrum; the theory is clarified, to obtain
the conditions for which, with incoherent light,  one gets  a redshift
similar to a Doppler shift rather than Raman lines.
The explanation of the appearance of the same
absorption line with various z becomes trivial, requiring. only a static
halo and a static magnetic field.

\end{abstract}


\section{Introduction}

Some astrophysicists think that Doppler effect cannot explain all
redshifts observed in astrophysics, but they were unable to find a
serious alternative theory. 
It is strange that they did not notice that redshift are observed
for a long time in the interaction of short pulses of light with matter
\cite{Giordmaine, Treacy}. Yan et al. \cite{Yan} gave a theory in a particular
case, but the conditions in which a redshift appears rather than raman
lines were not studied precisely.
Making the meaning of \textquotedblleft short pulse" clearer, we will apply the effect to usual
incoherent light, in particular to the light coming from the stars.
\section{ The space coherence of the scattered light}
It was tried to explain the redshift by a Raman scattering; the first reason
for which it does not work is : the Raman scattering is incoherent, that is
the scattered light is spread in all directions so that the images are blurred.

A laser is a source of space coherent light: a large source can emit for
instance a nearly plane wave. Light amplifiers are able to increase the intensity
of a light beam without any perturbation of the shape of the wave surfaces.

This property of light amplifiers is known in old optics: A small fraction
of the Rayleigh diffusion is incoherent, producing, for instance the blue of
the sky, but most Rayleigh diffusion is coherent, it produces a wave late of
$\pi/2$ which interferes with the incident beam, giving the refraction.

In the theory of refraction, the refracting molecules perform two-photons
virtual transitions, so that the initial state (say the ground state) is the
same than the final state. What happens if the ground state is split by a
perturbation ? If the perturbation is large, incoherent Raman appears, but
for which perturbation do we have the turn from space coherence to incoherence ?

The origin of the incoherence of ordinary Raman scattering is the collisions;
consider the molecules on a wave surface. In a classical scheme, they start
to vibrate all in phase (in a quantum scheme, the computations of scattering
are the same for all molecules); thus, all molecules emit wavelets similar to
Huygens wavelets which construct a wave surface identical to the incident wave
surface. These new waves may be Rayleigh, producing the refraction, or Raman.
It seems that Raman scattering is coherent! It is not because the collisions
stop the vibrations of the molecules; as the phase of the scattered wave is
late of $\pi/2$ when the oscillator restarts, this phase depends on the instant
of the collision. This is not important for Rayleigh scattering because the
difference of phase between the incident and scattered waves does not depend
on the time, but it makes the phases stochastic for Raman scattering, so that
this scattering is incoherent.

{\it Condition 1:

The Raman scattering is coherent if the duration of the light pulses is shorter
than the time between two collisions in the gas.}

However, in a gas, the dispersion generally changes the relative phases of
the incident and scattered beams, so that the phases of the beams which are
scattered on different wave surfaces are different: their interference is
constructive if the medium is thin; if it is thick, it becomes destructive;
the scattered intensity cannot be large \footnote {The scattered intensity
may be large on a cone if the beams are diffraction limited; we suppose here
that the beams are wide}.
\section{ Interference of the incident and scattered beams}
The second reason for which Raman scattering does not work is that the
considered Raman transitions let appear new lines. Successive scatterings do
not deplace the exciting line but diffuse it. 

This interference is trivial in the appearance of refraction through the
Rayleigh scattering. The definition of the frequency of a pulse is limited by
the length of the pulse; how is it possible to distinguish two frequencies ?
A simple computation shows that two sine functions may be added into a single
one, with a good precision if their phaseshift during the pulse is lower enough
than $\pi$. The final frequency is intermediate between the initial ones.
Thus we get

{\it Condition 2:

The appearance of Raman lines is replaced by a frequency shift if the half period
of the beats between the incident and scattered waves is shorter than the length
of the light pulses.}

More precisely, the mathematical equivalence between two waves of different
frequencies and a single wave of intermediate frequency leads to a vagueness
of the result; this vagueness is removed by the dispersion which destroys the
two frequencies solution.
\section{ Application to incoherent light}
Blackbody light is very incoherent, made of very short pulses. The light emitted
by excited atoms, or blackbody light filtered by an absorbing gas is made of much
longer pulses. Generally, these pulses are shorter than 10 nanoseconds, that is
an interferometer does not give fringes if its difference of path is larger than
3 meters. Thus, we have some orders of magnitude in the scale of time between
ordinary incoherent light and short pulses of light.

To fulfil condition 1 with incoherent light, the pressure of the gas must be very
low; to fulfil condition 2, the distance of the Raman levels must be lower than
50 MHz, that is the molecule must have an hyperfine structure.
Most molecules and atoms have an hyperfine structure; the exceptions are (roughly)
simple molecules with an even number of electrons and atoms with a small number of electrons.

This effect, very similar to refraction is very important; the computation of an
order of magnitude shows that if this effect produces the whole redshift of
nebulae, the necessary amount of active molecules $(H_2^+, NH_2, OH,...)$ is of the
order of 20 per cubic meter \cite{M1}.

The energy lost by the redshift is transferred, in a parametric process, to the
incoherent low energy (2.7K) radiation.

\section{ Application to the Ly$\alpha$ lines of quasars}
The standard explanation of the appearance of the same line with various z
requires thin clouds moving with very high speeds. We will suppose only the
existence of a static high temperature halo over an extremely high temperature
kernel, and the presence of a static magnetic field which falls to zero at 3 or
4 altitudes on the sight line \cite{M2}.

The atoms which have no hyperfine structure where the field is nearly zero, acquire
a Zeeman hyperfine structure elsewhere. When the light propagates near a zero of
the field, the absorption lines are written into the spectrum. Then, between two zeros,
the spectrum
is redshifted, but the absorption is permanently displaced in the spectrum, so
that the intensity of the spectrum decreases slightly, without an appearance of lines. Then,
near the following zero, the lines are written in a new place.

\section{ Conclusion}
It seems evident that the evolution of optical properties of light with the
pressure of the gases may be extrapolated to very low pressures. It is not exact
because the electromagnetic interactions between the molecules, in particular
the collisions, play an important role, and these interactions decrease at pressures
which are more commonly reached in the space than in the laboratory.


\section*{References}

\begin{thebibliography}{4}
\bibitem{Giordmaine} J. A. Giordmaine, M. A. Duguay \& J. W. Hansen, 1968,
{\it IEEE J. Quantum Electron.}, {\bf 4}, 252 
\bibitem{Treacy} E. B.Treacy, 1968, {\it Phys. Letters}, {\bf 28A}, 34 
\bibitem{Yan} Y.-X. Yan, E. B. Gamble Jr. \& K. A. Nelson, 1985,
{\it J. Chem Phys.}, {\bf 83}, 5391 
\bibitem{M1} J. Moret-Bailly, 1998, {\it Quantum Semiclass. Opt.}, {\bf 10},  L35 

\bibitem{M2} J. Moret-Bailly, 1998, {\it Ann. Phys. Fr.}, {\bf 23}, C1-235

\end{thebibliography}
\end{document}